\documentclass[aps,prb,twocolumn,showpacs,superscriptaddress]{revtex4}

\usepackage{amssymb}
\usepackage{amsmath}
\usepackage{graphicx}

\begin{document}

\title{Ideal Quantum non-demolishing measurement of a flux qubit at variable
bias}
\author{Ying-Dan Wang}
\affiliation{Department of Physics, University of Basel,
Klingelbergstrasse 82, 4056 Basel, Switzerland}
\author{Xiaobo Zhu}
\affiliation{NTT Basic Research Laboratories, NTT Corporation, 3-1,
Morinosato Wakamiya, Atsugi-shi, Kanagawa 243-0198, Japan}
\author{Christoph Bruder}
\affiliation{Department of Physics, University of Basel,
Klingelbergstrasse 82, 4056 Basel, Switzerland}

\begin{abstract}
We propose a scheme to realize a quantum non-demolition (QND)
measurement of a superconducting flux qubit by a Josephson bifurcation
amplifier. Our scheme can implement a perfect QND measurement for a
qubit subject to a variable magnetic bias.  Measurement back-action
induced qubit relaxation can be suppressed and hence the QND fidelity
is expected to be high over a wide range of bias conditions.
\end{abstract}

\maketitle


\section{Introduction}

Quantum non-demolition (QND) measurements enable repeated measurements
on quantum objects with accuracy levels exceeding the standard quantum
limit~\cite{Braginsky1996}. Such QND measurements on superconducting
flux qubits have been reported~\cite{Lupascu2007,Picot2010}. However,
these QND schemes work only far away from the degeneracy point (the
`sweet spot' where the sensitivity to noise is minimized), and the QND
criterion is only approximately satisfied. Here we propose an ideal
QND measurement scheme of a flux qubit that can be applied in a wide
range of bias conditions. The QND fidelity for this measurement is
expected to increase significantly as compared to previous proposals.

In quantum mechanics, measurements induce back-action to the system
under investigation due to Heisenberg uncertainty. This back-action
puts a fundamental limit on the precision of repeated measurements.
In order to beat the standard quantum
limit~\cite{Braginsky1980,Braginsky1996}, the concept of QND
measurement was developed in the context of gravitational wave
detection where repeated measurements beyond the standard quantum
limit are required~\cite{Bocko1996}. This concept has been extended
from gravitational wave detection to other physical systems. A number
of experiments has been performed in a micromechanical
system~\cite{Hertzberg2010} and quantum optical
systems~\cite{Grangier1998}.  This special type of measurement leaves
the output state unaffected by subsequent measurements and the free
evolution of the system. QND measurements are crucial to overcome
detector inefficiencies, and to quantify the external disturbance to
the QND variables. It is also found to have more versatile
applications such as error correction~\cite{Steane1996}, one-way
quantum computing~\cite{Duer2003}, low-noise
amplification~\cite{Levenson1993} and entanglement
generation~\cite{Helmer2009,Bishop2009}.

In superconducting qubit systems, weak continuous QND measurements
on superconducting transmon qubits have been realized in the
dispersive limit~\cite{Wallraff2004}. This circuit QED system has
also been used to detect single microwave photons in a coplanar wave
guide~\cite{Johnson2010}. Using the Josephson bifurcation amplifier,
strong projective QND measurements have been demonstrated for
quantronium qubits, flux qubits, and transmon
qubits~\cite{Boulant2007,Lupascu2007,Mallet2009}. In order to
implement QND detection for a continuous QND variable, a number of
criteria have to be satisfied~\cite{Braginsky1980,Bocko1996}. Among
them, the most restrictive one is that the system free Hamiltonian
$H_{s}$ commutes with the interaction $H_{int}$ between the system
and the detector, i.e., $[H_{s},H_{int}] =0$.  For existing flux
qubit measurements~\cite{Lupascu2007,Picot2010}, this QND criterion
is only approximately satisfied when the qubit is biased far away
from the degeneracy point. However, the quantum coherence times for
solid-state qubit vanish rapidly in this regime.  QND detection
close to the qubit degeneracy point is therefore desired.  Moreover,
to acquire full qubit control, the qubit bias has to be changed
during various operations. After implementing an operation at a certain
bias, it is desirable to be able to carry out a QND
measurement at that point, without adiabatically shifting back to
another bias value.  In this paper, by introducing an rf SQUID
coupler to mediate the interaction between a flux qubit and the
detector, a Josephson bifurcation amplifier (JBA), we find a
detection scheme that allows to implement a QND measurement at
arbitrary bias including the degeneracy point. Moreover, our scheme
works beyond the dispersive limit and can be extended to the case of
strong qubit-detector coupling. This will help to improve the
readout contrast to achieve a higher measurement fidelity and
shorter measurement times.

Another advantage of this scheme is the possibility to improve the
so-called QND fidelity, which quantifies the accuracy of repeated
measurements. In QND measurements by a
JBA~\cite{Lupascu2007,Boulant2007}, the drive on the JBA is first
ramped to the bifurcation point to induce transitions between two
bistable states. It is then reduced to maintain a latching plateau.
The circuit geometry in the previous experiments does not implement
an ideal QND measurement, i.e., $[H_{s},H_{int}] \neq 0$. Qubit
relaxation is then accelerated by the forced oscillations of the
nonlinear resonator. The population fraction lost during the
latching plateau and the preparation stage of the subsequent measurement
limit the accuracy of the subsequent measurement. It turns out that the
JBA induced qubit relaxation is the main limiting factor for the QND
fidelity~\cite{Picot2008}. In our design, if proper control on the
bias is acquired, the detection scheme is an ideal QND
measurement. The QND fidelity is only limited by environment-induced
qubit relaxation, which is usually one order of magnitude smaller than
the JBA induced relaxation.

The structure of this paper is as follows: In Sec.~II, we describe
the circuit layout and the effective mutual inductance between the
qubit and the JBA. The QND feature of this detection is analyzed in
Sec.~III, where two situations with bias at and off the degeneracy
point are discussed respectively. In Sec.~IV, we revisit the working
principle of the JBA and calculate the qubit relaxation rates in the
measurement process. With those relaxation rates, the fidelity of
the QND measurement is evaluated. Section~V discusses and summarizes
our results.

\section{The circuit layout}

\begin{figure}[tp]
\begin{center}
\includegraphics[bb=100 247 490 587,scale=0.5,clip]{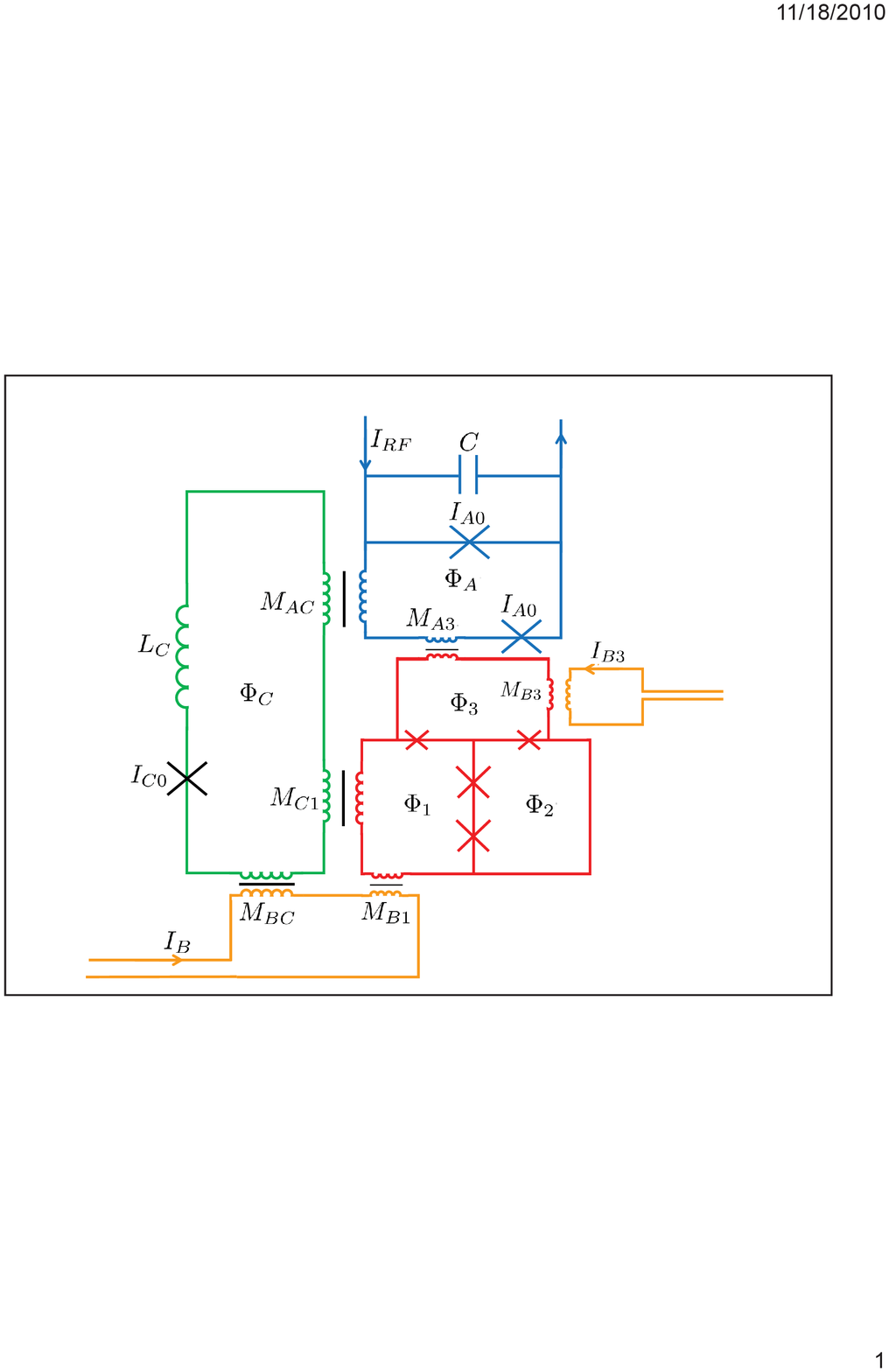}
\end{center}
\caption{(Color online) Schematic diagram of the circuit. Red
part: gradiometer flux qubit to be measured. Blue part: measurement
device, a Josephson bifurcation amplifier formed by
a dc SQUID shunted by a capacitance. Green part: rf SQUID
acting as a tunable coupler. Orange parts: bias circuits.}
\label{fig:circuit}
\end{figure}

Previous measurements can only work in the regime far away from the
degeneracy point. This is because the measurement circuit (e.g. a
Josephson bifurcation amplifier) can only be coupled to supercurrents in
the loop. However in the conventional 3-Josephson junction
design~\cite{Orlando1999,Mooij1999}, the current states are the
eigenstates of the system only if the qubit is biased to the
degeneracy point. A natural solution for this problem is to use a
gap-tunable qubit~\cite{Paauw2009,Fedorov2010,Zhu2010} and couple the
measurement device to the dc SQUID part. This will enable a QND
measurement when the qubit is biased at the degeneracy point.
However, to implement a QND measurement at variable bias, the coupling
with the measurement device has be to mediated in a way that it can
always follow the eigenstates of the system. In this paper, we
introduce a tunable coupler between the flux qubit and the Josephson
bifurcation amplifier. The qubit shares one control line with the
tunable coupler. As the qubit bias is varied, the qubit coupling to
the Josephson bifurcation amplifier is modified simultaneously. We
find that under certain conditions, a perfect QND measurement can be
performed at variable qubit bias.

The system we have in mind is shown is shown in
Fig.~\ref{fig:circuit}.  It is composed of four parts: the system to
be measured (red part), the measuring apparatus (blue part), the
coupler (green part) and the bias circuits (orange parts). The
system to be measured is a gradiometer-type superconducting flux
qubit~\cite{Paauw2009,Fedorov2010} which contains four Josephson
junctions in three loops: The two lower loops (the main qubit loops)
and the upper loop (the dc SQUID loop) penetrated by magnetic fluxes
$\Phi_{1}$, $\Phi_{2}$, and $\Phi_{3}$. The two junctions in the
dc SQUID loop are assumed to have identical Josephson energies
$\alpha _{0}E_{J}$, here $\alpha _{0}$ is the ratio between the
Josephson energy of the smaller junctions and that of the two bigger
junctions. The other two junctions are assumed to have the Josephson
energy $E_{J}$. The total Josephson energy of the circuit
is~\cite{Wang2010}
\begin{equation}
E_{J}\cos \varphi _{1}+E_{J}\cos \varphi _{2}+\alpha E_{J}\cos
\left( 2\pi \Phi _{t}/\Phi _{0}-(\varphi _{1}+\varphi _{2})\right)
\end{equation}%
where $\Phi _{t}\equiv (\Phi_{1}-\Phi_{2})/2$,
$\alpha=2\alpha_{0}\cos(\pi \Phi_{3}/\Phi_{0})$,
and $\varphi _{k}$ ($k=1,2,3,4$) is the phase difference across the
$k$-th Josephson junction. If $\Phi_{t}$ is chosen close to
$\Phi_{0}/2$ where $\Phi _{0}=h/(2e)$ is the flux quantum,
the circuit dynamics can be effectively described in a two-level
subspace of a double-well potential, and thus constitutes a flux
qubit~\cite{Mooij1999,Orlando1999}. Together with the charging energy,
the total Hamiltonian of the qubit is
\begin{equation}
H=\varepsilon (\Phi _{t})\sigma _{z}+\Delta (\Phi _{3})\sigma _{x}\, .
\end{equation}%
The Pauli matrices read $\sigma _{z}=|0\rangle \langle 0|-|1\rangle
\langle 1|$, $\sigma _{x}=|0\rangle \langle 1|+|1\rangle \langle
0|$, where $|0\rangle$ and $|1\rangle$ denote the states with
clockwise and counterclockwise currents in the outer loop. The
energy spacing of the two current states is $\varepsilon (\Phi
_{t})\equiv I_{p}(\Phi _{t}-\Phi _{0}/2) $, with $I_{p}$ the
magnitude of the classical persistent current in the loop. The
tunneling amplitude between the two states $\Delta (\Phi _{3})\equiv
\Delta (\alpha )$ depends on the bias in the dc SQUID loop. Note
that in contrast to the original flux qubit
design~\cite{Mooij1999,Orlando1999}, this gradiometer flux qubit is
insensitive to homogeneous fluctuations of the magnetic
flux~\cite{Paauw2009}. More importantly, it enables the JBA to
couple with the dc-SQUID loop without changing the total bias flux
of the qubit.

The detector for the flux qubit is a Josephson bifurcation amplifier
(JBA)~\cite{Siddiqi2004} (blue part in Fig.~\ref{fig:circuit}),
which in our scheme is a dc SQUID shunted by a capacitance $C$,
subject to a microwave drive $I_{RF}\cos(\omega_{d}t+\phi_A)$. The
JBA SQUID loop contains two identical Josephson junctions of
critical current $I_{A0}$. The phase differences across the two
junctions are denoted by $\varphi _{A1}$, $\varphi _{A2}$
respectively. The current in the loop is $I_{A}=\bar{I}_{A}(\Phi
_{A}) \cos \varphi_A$, with $\Phi_{A}$, the flux bias in the JBA
SQUID, set by external coils, $\bar{I}_{A}(\Phi _{A}) =2I_{A0}\sin (
\pi \Phi _{A}/\Phi _{0})$, and $\varphi_A =(\varphi_{A1}+\varphi
_{A2})/2$. The JBA circuit forms a driven nonlinear resonator which
exhibits bistable behavior with hysteresis. With appropriate drive
sequences, a transition to one of the bistable states is correlated
with the qubit states in a probabilistic way. Therefore the qubit
state can be read out by the phase of the transmitted or reflected
microwave.

The flux qubit is coupled to the JBA through their mutual
inductance. There are two contributions to their mutual inductance:
the direct mutual inductance (DMI) $M_{Ak}$ ($k =1$, $2$, and $3$
denotes the different loops in the qubit) and the effective mutual
inductance (EMI) $M_{Ak}^{\prime}$. Hence the JBA produces flux
biases to the qubit loops of the form $(M_{Ak}+M_{Ak}^{\prime})
I_{A}$. The EMI is induced by the nearby rf SQUID which acts as a
coupler for the qubit and the JBA. The self-inductance of the
coupler is assumed to be much larger than the mutual inductances and
the dynamics of the coupler is confined to its lowest energy
bands~\cite{Averin2003,Maassen2005}. The DMI is fixed by fabrication
processes while the EMI is tunable by the magnetic bias of the
coupler $\Phi _{C}$ as~\cite{Maassen2005,Allman2010}
\begin{equation}
M_{Ak}^{\prime}(\Phi_{C}) =-\frac{M_{AC}M_{Ck}}{L_{C}}
\frac{\beta_{c}\cos\theta_{0}}{1+\beta_{c}\cos \theta_{0}}\, ,
\label{mui}
\end{equation}%
where $\theta_{0}$ satisfies the nonlinear equation
\begin{equation}
\theta_{0}=\frac{2\pi}{\Phi_{0}}(\Phi_{C}+M_{AC}I_{A0})
-\beta_{c}\sin \theta_{0}
\end{equation}%
with $\beta_{c}=2\pi L_{C}I_{C0}/\Phi_{0}$, $L_{C}$ is the
self-inductance of the coupler, and $I_{C0}$ the circulating coupler
current. In particular, if the coupler is biased at
$\Phi_{C}=((2n+1/2)\pi +\beta_{c}) \Phi _{0}/2\pi -M_{AC}I_{A0}$ or
$((2n+3/2)\pi -\beta_{c}) \Phi_{0}/2\pi -M_{AC}I_{A0}$ ($n$ is an
arbitrary integer), the effective mutual inductance vanishes, $M_{Ak
}^{\prime }=0$. Thus, for this bias condition, the EMIs between the
JBA and all the qubit loops are canceled, and only the DMIs
contribute to the coupling.

Besides tunability, there is another important difference between
the DMI and the EMI: the DMI is symmetric with respect to the qubit
loops 1 and 2, while the EMI is not symmetrical, that is,
$M_{A1}=M_{A2}$, while $M_{A1}^{\prime}\neq M_{A2}^{\prime }$
(since $M_{C1}\neq M_{C2}$). Hence only the EMI couples the JBA to
the gradiometer qubit flux $\Phi_{t}$ in the form
\begin{equation}
\Phi_{t}=(M_{A1}^{\prime}-M_{A2}^{\prime}) I_{A}\, .
\label{EMI_gradiometer}
\end{equation}

In our scheme, the whole chip is biased by external coils so that a
homogeneous magnetic field threads all the loops. By choosing the area
of each loop appropriately, the required background bias values can be
imposed. Besides the coupling to the external coils, the dc SQUID loop
of the qubit is also coupled with an on-chip bias current $I_{B3}$
through $M_{B3}$.  The qubit loop 1 shares another on-chip bias (the
lower orange part) with the coupler. A bias current $I_{B}$ in this
bias line couples to the qubit loops and the coupler loop through
mutual inductances $M_{Bk}$ and $M_{BC}$. In the following discussion,
we will see that this shared bias is crucial for the possibility to do
a QND measurement at arbitrary bias.

\section{QND nature of the detection scheme}

In order to analyze the QND nature of the detection scheme, we first look at
the situation that the bias is set at the degeneracy point and then study
the case of a general (off-degeneracy) bias.

\subsection{Degeneracy point}

We first look at the case when the qubit is biased at the degeneracy
point $\Phi_{tb}=\Phi_{0}/2$. At this point, the first-order flux
noise disappears so that the qubit quantum coherence can be
preserved longer.

The qubit is biased at the degeneracy point by trapping one flux
quantum~\cite{Paauw2009,Fedorov2010}. The bias current is set to
zero $I_{B}=0$ and the flux bias of the coupler is set by external
coils to be
\begin{equation}
\Phi_{Cb}=\frac{\Phi _{0}}{2\pi }\left( \frac{\pi }{2}+
\beta_{c}\right) -M_{AC}\bar{I}_{A}\, .
\end{equation}
According to the discussion following Eq.~(\ref{mui}), the
effective mutual inductance $M_{Ak}^{\prime}$ vanishes at this
bias. Thus, the qubit only couples to the JBA through the direct
mutual inductance. As shown in Eq.~(\ref{EMI_gradiometer}), this
means the JBA has no influence on $\Phi_{t}$, but only couples to
$\Phi_{3}$. If $\pi M_{A3}\bar{I}_{A}\ll \Phi _{0}$, the Hamiltonian
can be expanded to first order as~\cite{Wang2010}
\begin{equation}
H_{q}=H_{q0}+H_{I}\, ,
\label{H1}
\end{equation}%
where $H_{q0}$ is the free Hamiltonian of the qubit
\begin{equation}
H_{q0}=\Delta (\Phi_{3b})\sigma _{x}\, ,
\end{equation}%
and $H_{I}$ is the interaction between the qubit and the JBA
\begin{equation}
H_{I}=\lambda(\Phi_{3b}) \sigma _{x}\cos \varphi_A\, . \label{ham_i}
\end{equation}%
where $\Phi_{3b}$ is the total flux bias of the dc SQUID loop
(generated by both external coils and $I_{B3}$). The coupling
coefficient is
\begin{equation}
\lambda(\Phi_{3b}) =-\frac{\pi M_{A3}\bar{I}_{A}}{\Phi
_{0}}\kappa(\Phi_{3b})
\end{equation}%
with%
\begin{equation*}
\kappa(\Phi_{3b}) =2\alpha_{0}\sin\left(\pi \frac{\Phi_{3b} }{\Phi
_{0}}\right) \left. \frac{d\Delta(\alpha)}{d\alpha }
\right\vert_{\alpha =\bar{\alpha}}
\end{equation*}%
and $\bar{\alpha}=2\alpha_{0}\cos(\pi\Phi_{3b}/\Phi_{0})$. The
coupling energy between the qubit and the JBA is tunable by
$\Phi_{3b}$.

Equations~(\ref{H1}) -- (\ref{ham_i}) show that the free Hamiltonian
commutes with the interaction Hamiltonian. If one chooses $\sigma_{x}$
as the measurement observable, a QND measurement can be implemented.

\subsection{General (off-degeneracy) bias}

If we change the current in the shared bias by a small amount
$I_{B}=\delta I_{B}$, the qubit is biased away from the degeneracy
point, and the corresponding bias change in the qubit loop is
$\delta \Phi _{t}=(M_{B1}-M_{B2})\delta I_{B}$. Since the coupler
shares the same bias, the magnetic flux penetrating the coupler bias
is also shifted by a small amount $M_{BC}\delta I_{B}$
($M_{BC}\delta I_{B}\ll {\Phi}_{Cb}$ is always satisfied in the
relevant operation regime). This will induce a non-zero effective
mutual inductance $ M_{Ak }^{\prime }({\Phi}_{Cb}+M_{BC}\delta
I_{B}) =-(2\pi/\Phi_{0})^{2}I_{c0}M_{AC}M_{Ck }M_{BC}\delta I_{B}$.
As we discussed after Eq.~(\ref{EMI_gradiometer}), a non-zero EMI
will couple the JBA to the qubit flux bias $\Phi_{t}$ as well as
$\Phi_{3}$.  The qubit Hamiltonian under this bias reads
\begin{equation}
H_{q}=H_{q0}+H_{I}
\end{equation}%
with%
\begin{equation}
H_{q0}=\Omega _{z}\sigma _{z}+\Omega _{x}\sigma _{x}
\end{equation}%
and the interaction Hamiltonian%
\begin{equation}
H_{I}=\left(\lambda_{z}\sigma_{z}+\lambda_{x}(\Phi_{3b})
\sigma_{x}\right) \cos \varphi_A
\end{equation}%
with $\Omega _{z}=I_{p}\delta \Phi _{t}$, $\Omega _{x}=\Delta
(\Phi_{3b})$
and%
\begin{equation}
\lambda_{z}=I_{p}(M_{A1}^{\prime }-M_{A2}^{\prime}) \bar{I}_{A}\, ,
\end{equation}%
\begin{equation}
\lambda_{x}(\Phi_{3b}) ={\lambda}(\Phi_{3b}) \left(
1+\frac{M_{A3}^{\prime }}{M_{A3}}\right)\, .
\end{equation}

If we define a parameter
$\eta =\lambda_{z}\Omega_{x}/\lambda_{x}\Omega_{z}$,
it is straightforward to see that the free qubit Hamiltonian commutes
with the interaction Hamiltonian when $\eta =1$. Therefore, a sufficient
condition to implement a QND measurement at variable flux bias is
\begin{equation}
\eta(\Phi_{3b}) =4\pi \frac{M_{BC}I_{c0}}{\Phi
_{0}}\frac{M_{C1}}{M_{B1}}\frac{M_{AC}}{M_{A3}}\frac{\Delta
(\Phi_{3b})}{\kappa(\Phi_{3b})}=1 \label{condition}
\end{equation}%
where we have neglected $M_{B2}$, $M_{C2}$, and $M_{C3}$ since they are much
smaller than the other mutual inductances.

In other words, if $\eta =1$, as the qubit is biased away from the
degeneracy point, the interaction with the JBA is changed
accordingly, so that the interaction Hamiltonian always commutes
with the qubit free Hamiltonian. This condition is possible to be
satisfied experimentally, e.g., if $\alpha_{0}=0.4$, and the bias
$\Phi_{3b}$ satisfies $\bar{\alpha}=0.7$.  At this bias,
$\Delta/(d\Delta /d\bar{\alpha})\approx -0.11$~\cite{Wang2009}. If
$I_{C0}=1$~$\mu$A, $M_{BC}=23.5$~pH, $M_{C1}=25$~pH, $M_{B1}=5$~pH,
$M_{AC}=25$~pH, $M_{A3}=5$ pH, then $\eta =1$. Note that $\eta$
depends on the bias $\Phi_{3b}$ which is tunable \emph{in situ},
i.e., by tuning $I_{B3}$. Thus, errors in the fabrication process
can be compensated to satisfy Eq.~(\ref{condition}), the condition
for QND detection. Note that $\Phi_{3b}$ is determined by this
condition since all the other parameters are fixed by fabrication.
Therefore, this QND scheme works for variable bias values of the
main qubit loop, but it does not work for variable bias values of
the dc SQUID loop in the general case. However, we would also like
to point out that if the qubit is biased away from the degeneracy
point, by changing $\varepsilon$, any arbitrary single-qubit
operation can be implemented; in this sense, it is not necessary to
tune $\Phi_{3b}$. Also, if the qubit is biased at the degeneracy
point, which is also the situation in which tuning $\Phi_{3b}$ is
meaningful, the QND measurement can be implemented for variable
$\Phi_{3b}$.

For $\eta =1$, the results obtained for the degeneracy point and
general (off-degeneracy) bias can be written in a uniform way as
\begin{equation}
H_{q0}=\Omega \tilde{\sigma}_{z}
\end{equation}%
with $\Omega =\sqrt{\Omega _{x}^{2}+\Omega _{z}^{2}}$, and
$\tilde{\sigma}_{z}=(\Omega _{z}\sigma _{z}+\Omega _{x}\sigma
_{x})/{\Omega }$. The interaction Hamiltonian is
\begin{equation}
H_{I}=\lambda \tilde{\sigma}_{z}\cos \varphi_A
\label{int}
\end{equation}%
with $\lambda =\sqrt{\lambda _{z}^{2}+\lambda _{x}^{2}}$. At the
degeneracy point, $\Omega _{z}=\lambda _{z}=0$, so that
$\tilde{\sigma}_{z}=\sigma _{x}$.

\section{Measurement fidelity}

The JBA is an oscillator with nonlinear Josephson inductance. Under
a strong microwave drive, the Josephson energy of the junction
$-E_{JA}\cos \varphi_A $ is expanded beyond the harmonic approximation and
the classical dynamics can be described by a Duffing
oscillator~\cite{Landau_mechanics}. For a certain range of drive
conditions, the nonlinear oscillator exhibits bistable behavior with
hysteresis~\cite{Landau_mechanics,Siddiqi2004}. The two possible
stable states correspond to different oscillation amplitudes and
phases, which can be distinguished by transmitted or reflected
microwave signals~\cite{Boulant2007,Lupascu2007,Mallet2009}.
Switching between the two stable states happens when the drive power
reaches a certain threshold. The switching probability depends on the
value of the nonlinear inductance, which in our case depends on the
states of the qubit through the mutual inductance. This is because the
effective Josephson energy of the junctions of the JBA is modified
by the interaction Eq.~(\ref{int}) as
$E_{JA}(\tilde{\sigma}_{z}) =\bar{I}_{A}\Phi _{0}/2\pi -\lambda
\tilde{\sigma}_{z}$. Therefore, measuring the phase of the
transmitted microwave signal, one can read out the state of the qubit.

The back-action from the measurement device destroys the phase
coherence of the qubit states during the read-out process.  Besides
dephasing, the back-action could also induce relaxation to the
qubit. This is the case for a qubit Hamiltonian
$H_{q0}=\Omega \tilde{\sigma}_{z}+\tilde{\Delta} \tilde{\sigma}_{x}$
with a small non-ideal QND fraction $\tilde{\Delta}$, see e.g.
the QND measurement of Ref.~\onlinecite{Lupascu2007} where
$\tilde{\Delta}/\Omega \approx 0.34$.
The JBA is strongly coupled to a dissipative environment while
weakly coupled to the qubit. Hence it serves as a bath for the
qubit. According to Eq.~(\ref{int}), the influence of the JBA on the
qubit can be described by its correlation function
\begin{equation}
G(\omega) =\int_{0}^{\infty }dte^{i\omega t}\langle \cos
\varphi_A(t) \cos \varphi_A(0)\rangle\, ,
\end{equation}%
and the induced decay rate can be calculated through the Fermi golden
rule. The Bloch Redfield rates induced by the operation of the JBA
are
\begin{eqnarray}
\Gamma _{r} &= &\frac{\lambda ^{2}\tilde{\Delta}}{\sqrt{\Omega
^{2}+\tilde{\Delta}^{2}}}\Re (G(\sqrt{\Omega^2+\tilde{\Delta}^2}))  \notag \\
\Gamma _{\varphi } &= &\frac{\lambda ^{2}\Omega }{\sqrt{\Omega ^{2}+
\tilde{\Delta}^{2}}}\Re (G(0))\, , \label{inde}
\end{eqnarray}%
where $\Gamma_{r}$ is the induced relaxation rate and
$\Gamma_{\varphi}$ is the induced dephasing rate. When the JBA is
ramped to the measurement level and the latching plateau, the
correlation function is prominently increased due to quantum
activation~\cite{Dykman2007}. Qubit decay is enhanced by the
measurement operation~\cite{Picot2008,Serban2010}. This results in
qubit relaxation and the measurement is driven away from the QND
regime. This induced relaxation has been found to be the main source
of measurement error~\cite{Lupascu2007,Picot2008}. One way to reduce
this back-action is working in the dispersive
limit~\cite{Mallet2009}. In our case, an ideal QND measurement is
possible, i.e. $\tilde{\Delta}=0$, so that $\Gamma _{r}=0$, i.e., the
JBA does not induce extra relaxation but only dephasing to the
qubit. Hence the QND condition can be preserved better in our
scheme and the QND fidelity can be improved.

Besides the induced decay rates Eq.~(\ref{inde}), there is another
decay mechanism due to the flux fluctuations of the environment.
This will perturb the fluxes in the qubit loops as $\delta \Phi
_{t}=\mu _{t}X$ and $\delta \Phi _{3}=\mu _{3}X$, where $X$
represents an environmental operator (such as a two-level
fluctuator) and $\mu _{t}$ ($\mu _{3}$) characterizes its coupling
strength to the different qubit loops. Hence the qubit is coupled to
the environment as $\Delta H=\xi _{t}X\sigma _{z}+\xi _{3}X\sigma
_{x}$, with $ \xi _{t}=I_{p}\Phi _{0}\mu _{t}$ and $\xi _{3}=-\pi
\mu _{3}\kappa(\Phi_{3b}) /\Phi _{0}$. In the interaction picture
\begin{eqnarray}
\Delta H_{I} &=&(\xi_{3}\sin\chi +\xi_{t}\cos\chi)
X(t) \sigma _{z}^{\prime}  \\
&&+(\xi_{3}\cos\chi -\xi_{t}\sin\chi) X(t)
\left( \sigma _{+}^{\prime }e^{i\Omega t}+\sigma _{-}^{\prime }e^{-i\Omega
t}\right)\, ,\notag
\end{eqnarray}%
with $\cos \chi =\varepsilon /\Omega $ and $\sin \chi =\Delta /\Omega $.

According to the Fermi golden rule, the relaxation rate is
$\Gamma_{\downarrow ,\uparrow }=(\xi _{3}\cos \chi -\xi _{t}\sin
\chi )^{2}S_{X}(\omega =\pm \Omega)$, where $S_{X}(\omega) =
\int_{-\infty}^{\infty }d\tau \langle X(\tau) X(0) \rangle
e^{i\omega \tau }$ is the flux noise spectrum. In a real experiment,
the flux noise could have multiple sources, such as two-level
fluctuators inside the barrier, high-frequency noise from the
control lines~\cite{Paauw2009}, and others.  Therefore the noise
spectrum may exhibit a complicated frequency dependence and have a
strong sample dependence.  In our discussion, we assume an Ohmic
noise spectrum ($f$-noise) for the environment bath plus a few peaks
due to two-level fluctuators
\begin{equation}
S_{X}(\omega) =\omega R_{0}\left( \coth \left(
\frac{\omega }{2k_{B}T_{0}}\right) +1\right)+\sum_i S_i
\delta(\omega-\omega_i)\, ,
\end{equation}
where $R_0$ is the Ohmic impedance and $T_0$ is the environmental
temperature. The QND fidelity of two successive measurements is
\begin{equation}
F_{\text{QND}}(\tau) =\frac{p(e|e) + p(g|g)}{2}=
\frac{\exp(-\Gamma _{\downarrow }\tau)
+\exp(-\Gamma _{\uparrow }\tau) }{2}
\label{fide}
\end{equation}
where $p(e|e)$ ($p(g|g)$) is the probability that the qubit state
$\left\vert e\right\rangle $ ($\left\vert g\right\rangle $) is
unchanged after the first measurement and $\tau$ is the time interval
between the two measurements.

\begin{figure}[tp]
\begin{center}
\includegraphics[width = 0.47\textwidth]{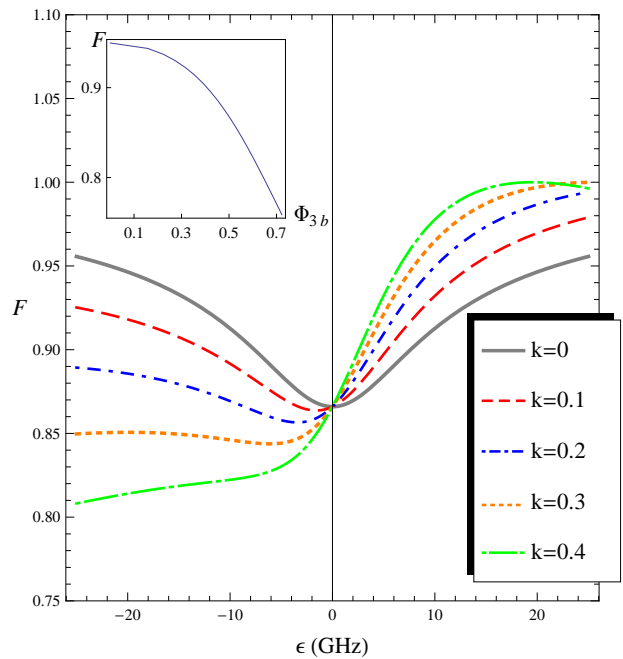}
\end{center}
\caption{(Color online) Dependence of the QND fidelity defined in
Eq.~(\ref{fide}) on the bias of the flux qubit for different values
of $k=\xi_3/\xi_t$. Inset: dependence of the QND fidelity on the
bias flux $\Phi_{3b}$ (in units of $\Phi_0/\pi$) in the dc SQUID
loop, when the qubit is biased at the degeneracy point.}
\label{fig:fidelity1}
\end{figure}
Figure~\ref{fig:fidelity1} shows the dependence of the QND fidelity
on the qubit bias $\varepsilon$ for different values of
$k=\xi_{3}/\xi_{t}$ (usually $k<1$ since the perturbation on the
main qubit loop in general has a larger influence than the
perturbation on the dc SQUID loop~\cite{Wang2009}). Here we assume
that the time interval between two measurements is $\tau =50$ ns,
$T_0=20$~mK, the qubit relaxation time is $250$~ns at the degeneracy
point, and the value of $\Delta $ is fixed at $7.8$ GHz. The plot
shows that the measurement fidelity remains rather high for a wide
range of bias values. Even at the degeneracy point $\varepsilon =0$
where the relaxation is strong, a measurement fidelity larger than
90\% can be achieved. As the bias is increased to the positive side,
the fidelity increases as the relaxation decreases. Far above the
degeneracy point, the measurement fidelity is very close to 100\%.
Note that the fidelity is not symmetrical with respect to the axis
$\varepsilon =0$ but becomes more symmetrical as $k$ decreases. At
$k=0$, the curve shows complete symmetry because noise only
contributes to the main qubit loop, it is symmetrical with respect
of the sign of the qubit bias. The inset of Fig.~\ref{fig:fidelity1}
shows that the QND fidelity at the degeneracy point decreases with
the SQUID bias $\pi\Phi_{3b}/\Phi_0$. This is because at the
degeneracy point, the qubit relaxation rate due to $f$-noise
increases linearly with the gap $\Delta$ and $\Delta$ increases with
the SQUID bias~\cite{Wang2009}.

The measurement fidelity can be used as a noise spectrometer
for environmental fluctuations. This is actually one of the main
applications of QND measurements: detecting perturbations to the
system. The QND nature of the measurement guarantees that the readout
back-action will not change the value of the observable. The
measurement fidelity therefore reflects the noise spectrum of the
environment. For example, the existence of one two-level fluctuator
inside the barrier~\cite{Lupascu2009,Kemp2010} would be revealed by a
corresponding peak in the QND fidelity at a certain bias.

\section{Discussion and Summary}
\begin{figure}[tp]
\begin{center}
\includegraphics[bb=100 247 490 587,scale=0.5,clip]{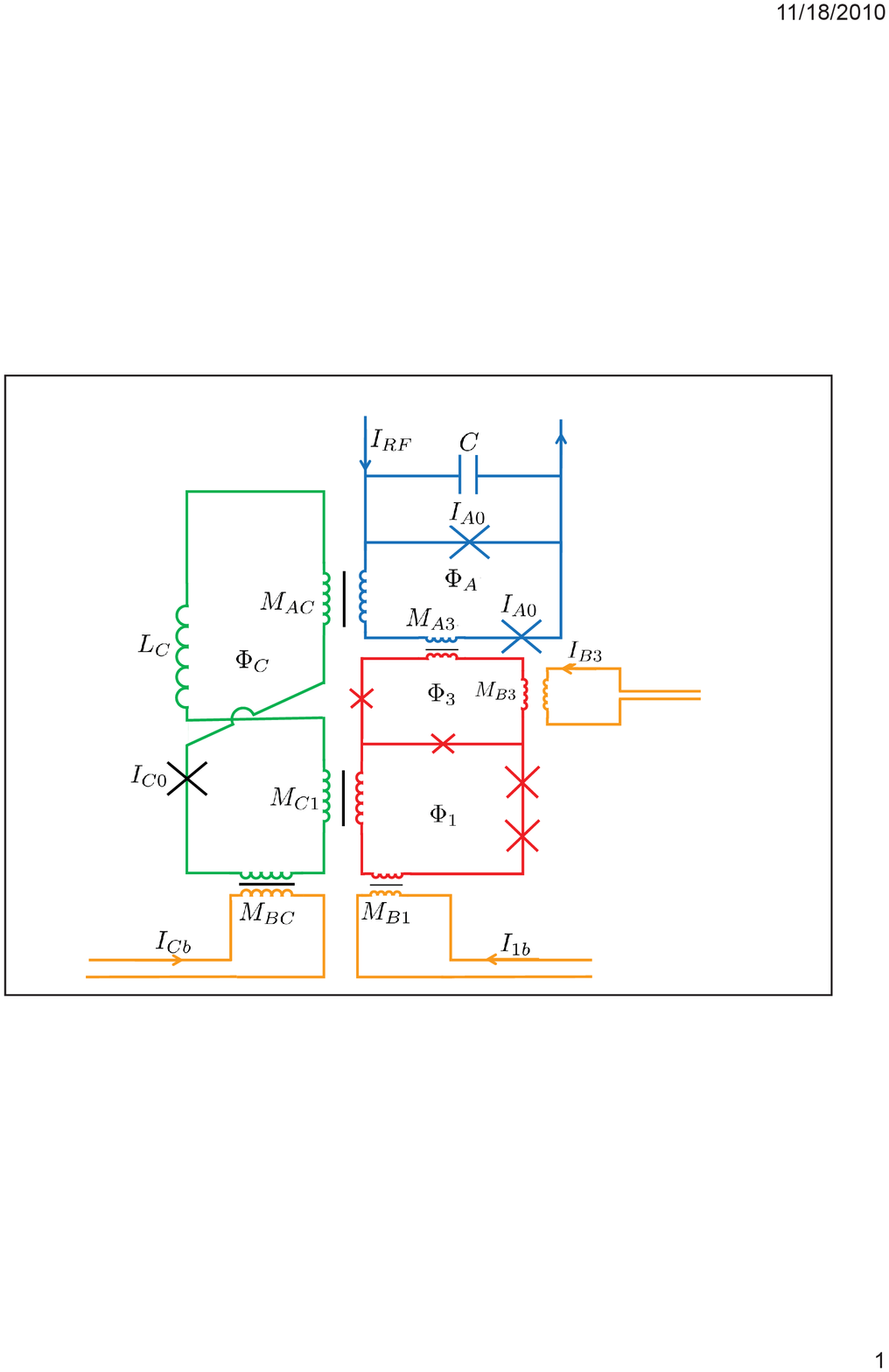}
\end{center}
\caption{(Color online) Schematic diagram of the circuit for a
non-gradiometer qubit. Red part: non-gradiometer flux qubit to be
measured. Blue part: measurement device, a Josephson bifurcation
amplifier formed by a dc SQUID shunted by a capacitance. Green part:
rf SQUID acting as a tunable coupler. Orange parts: bias circuits.}
\label{fig:circuit_ng}
\end{figure}

All the discussion above is based on the gradiometer type-flux
qubit. However, with a few modifications as explained below, the
measurement protocol can be adapted to non-gradiometer type flux
qubits with a tunable gap~\cite{Zhu2010}, see
Fig.~\ref{fig:circuit_ng}. Two current bias lines
$I_{Cb}$ and $I_{1b}$ are used to control the coupler and the qubit
separately in order to guarantee a QND measurement for a
non-gradiometer flux qubit at the degeneracy point.

\begin{figure}[tp]
\begin{center}
\includegraphics[bb=0 0 244 185,scale=0.9, clip]{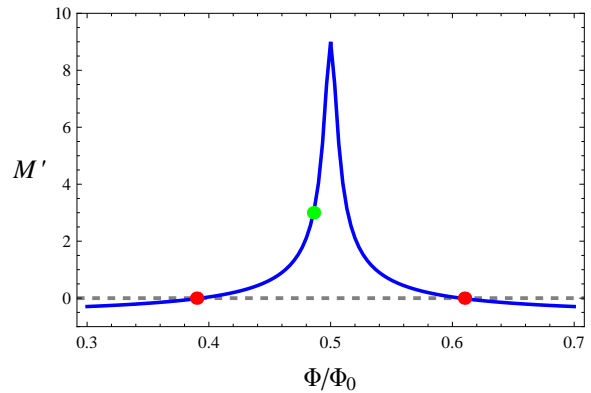}
\end{center}
\caption{(Color online) Dependence of the scaled effective mutual
inductance $M'=M'_{Ak}/(M_{AC}M_{Ck}/L_C)$ on the coupler bias
$\Phi=\Phi_{CB}+M_{AC}I_{A0}$, see Eq.~(\ref{mui}). The red dots on the two
sides (intersections with the dashed line) indicate the background bias in
the case of the gradiometer qubit, while the green point in the middle
is an example of the background bias for a non-gradiometer qubit. Here
$\beta_c=0.9$.}
\label{fig:effm}
\end{figure}

The background bias of a non-gradiometer qubit is sensitive to
homogeneous magnetic field fluctuations, but can be implemented
easily by external coils (while the gradiometer qubit requires the
technique to trap fluxoids). Also, it is possible to achieve a more
sensitive tuning comparing with the gradiometer qubit. This can be
seen from Fig.~\ref{fig:effm} which shows the scaled effective
mutual inductance with respect to the coupler bias. To achieve a
more sensitive tuning within the tunable range of the on-chip bias
(typically on the order of m$\Phi_0\equiv 10^{-3}\Phi_0$), the
coupler is desired to be pre-biased close to $\Phi_0/2$. However, in
the case of the gradiometer qubit, since $M'_{Ak}$ should be zero
when the qubit is biased at the degeneracy point, the background
bias should be set around the red points in Fig.~\ref{fig:effm},
i.e., relatively far from $\Phi_0/2$. For non-gradiometer qubits,
the background bias point is determined by the fabrication process.
If the mutual inductance between the JBA and qubit is large and the
coupler loop is twisted as indicated in Fig.~\ref{fig:effm}, the
background bias can be set closer to $\Phi_0/2$ (say, the green
point in Fig.~\ref{fig:effm}). As a result, a more sensitive tuning
can be achieved. For example, if we assume $M_{A3}=5$~pH,
$M_{A1}=0.5$~pH, $M_{AC}=10$~pH, $L_C=100$~pH, $M_{C1}=10$~pH,
$\bar{I}_A=1$~$\mu$A, for a change in the qubit bias
$\delta\Phi_t=2$~m$\Phi_0$, the coupler bias $\Phi_{CB}$ should be
tuned by $15$~m$\Phi_0$ in the case of the gradiometer qubit, while
only $0.4$~m$\Phi_0$ in the case of a non-gradiometer qubit.

In summary, we have studied a scheme to realize a quantum
non-demolition (QND) measurement for gradiometer-type flux qubits by a
Josephson bifurcation amplifier. We have shown that a perfect QND
measurement can be implemented for a qubit with variable magnetic
bias. The QND fidelity of this measurement is expected to be high over
a wide range of bias conditions. We have also discussed how to
generalize our scheme to non-gradiometer qubits. Our estimates
indicate that such a QND measurement may be realized experimentally,
and we hope that this will happen in the close future.

\section{Acknowledgment}

This work was financially supported by the EC IST-FET project SOLID,
the Swiss SNF, and the NCCR Nanoscience. Xiaobo Zhu was supported in
part by the Funding Program for World-Leading Innovative R\&D on Science
and Technology(FIRST), and KAKENHI Nos. 18001002 and 22241025 by
JSPS.


\begin{thebibliography}{33}
\expandafter\ifx\csname
natexlab\endcsname\relax\def\natexlab#1{#1}\fi
\expandafter\ifx\csname bibnamefont\endcsname\relax
  \def\bibnamefont#1{#1}\fi
\expandafter\ifx\csname bibfnamefont\endcsname\relax
  \def\bibfnamefont#1{#1}\fi
\expandafter\ifx\csname citenamefont\endcsname\relax
  \def\citenamefont#1{#1}\fi
\expandafter\ifx\csname url\endcsname\relax
  \def\url#1{\texttt{#1}}\fi
\expandafter\ifx\csname urlprefix\endcsname\relax\def\urlprefix{URL
}\fi \providecommand{\bibinfo}[2]{#2}
\providecommand{\eprint}[2][]{\url{#2}}

\bibitem[{\citenamefont{Braginsky and Khalili}(1996)}]{Braginsky1996}
\bibinfo{author}{\bibfnamefont{V.~B.} \bibnamefont{Braginsky}}
  \bibnamefont{and} \bibinfo{author}{\bibfnamefont{F.~Y.}
  \bibnamefont{Khalili}}, \bibinfo{journal}{Rev. Mod. Phys.}
  \textbf{\bibinfo{volume}{68}}, \bibinfo{pages}{1} (\bibinfo{year}{1996}).

\bibitem[{\citenamefont{Lupascu et~al.}(2007)\citenamefont{Lupascu, Saito,
  Picot, Groot, Harmans, and Mooij}}]{Lupascu2007}
\bibinfo{author}{\bibfnamefont{A.}~\bibnamefont{Lupascu}},
  \bibinfo{author}{\bibfnamefont{S.}~\bibnamefont{Saito}},
  \bibinfo{author}{\bibfnamefont{T.}~\bibnamefont{Picot}},
  \bibinfo{author}{\bibfnamefont{P.~C.~D.} \bibnamefont{Groot}},
  \bibinfo{author}{\bibfnamefont{C.~J. P.~M.} \bibnamefont{Harmans}},
  \bibnamefont{and} \bibinfo{author}{\bibfnamefont{J.~E.} \bibnamefont{Mooij}},
  \bibinfo{journal}{Nature Physics} \textbf{\bibinfo{volume}{3}},
  \bibinfo{pages}{119} (\bibinfo{year}{2007}).

\bibitem[{\citenamefont{Picot et~al.}(2010)\citenamefont{Picot, Schouten,
  Harmans, and Mooij}}]{Picot2010}
\bibinfo{author}{\bibfnamefont{T.}~\bibnamefont{Picot}},
  \bibinfo{author}{\bibfnamefont{R.}~\bibnamefont{Schouten}},
  \bibinfo{author}{\bibfnamefont{C.~J. P.~M.} \bibnamefont{Harmans}},
  \bibnamefont{and} \bibinfo{author}{\bibfnamefont{J.~E.} \bibnamefont{Mooij}},
  \bibinfo{journal}{Phys. Rev. Lett.} \textbf{\bibinfo{volume}{105}},
  \bibinfo{pages}{040506} (\bibinfo{year}{2010}).

\bibitem[{\citenamefont{Braginsky et~al.}(1980)\citenamefont{Braginsky,
  Vorontsov, and Thorne}}]{Braginsky1980}
\bibinfo{author}{\bibfnamefont{V.~B.} \bibnamefont{Braginsky}},
  \bibinfo{author}{\bibfnamefont{Y.~I.} \bibnamefont{Vorontsov}},
  \bibnamefont{and} \bibinfo{author}{\bibfnamefont{K.~S.}
  \bibnamefont{Thorne}}, \bibinfo{journal}{Science}
  \textbf{\bibinfo{volume}{209}}, \bibinfo{pages}{547} (\bibinfo{year}{1980}).

\bibitem[{\citenamefont{Bocko and Onofrio}(1996)}]{Bocko1996}
\bibinfo{author}{\bibfnamefont{M.~F.} \bibnamefont{Bocko}} \bibnamefont{and}
  \bibinfo{author}{\bibfnamefont{R.}~\bibnamefont{Onofrio}},
  \bibinfo{journal}{Rev. Mod. Phys.} \textbf{\bibinfo{volume}{68}},
  \bibinfo{pages}{755} (\bibinfo{year}{1996}).

\bibitem[{\citenamefont{Hertzberg et~al.}(2010)\citenamefont{Hertzberg,
  Rocheleau, Ndukum, Savva, Clerk, and Schwab}}]{Hertzberg2010}
\bibinfo{author}{\bibfnamefont{J.~B.} \bibnamefont{Hertzberg}},
  \bibinfo{author}{\bibfnamefont{T.}~\bibnamefont{Rocheleau}},
  \bibinfo{author}{\bibfnamefont{T.}~\bibnamefont{Ndukum}},
  \bibinfo{author}{\bibfnamefont{M.}~\bibnamefont{Savva}},
  \bibinfo{author}{\bibfnamefont{A.~A.} \bibnamefont{Clerk}}, \bibnamefont{and}
  \bibinfo{author}{\bibfnamefont{K.~C.} \bibnamefont{Schwab}},
  \bibinfo{journal}{Nature Physics} \textbf{\bibinfo{volume}{6}},
  \bibinfo{pages}{213} (\bibinfo{year}{2010}).

\bibitem[{\citenamefont{Grangier et~al.}(1998)\citenamefont{Grangier, Levenson,
  and Poizat}}]{Grangier1998}
\bibinfo{author}{\bibfnamefont{P.}~\bibnamefont{Grangier}},
  \bibinfo{author}{\bibfnamefont{J.~A.} \bibnamefont{Levenson}},
  \bibnamefont{and} \bibinfo{author}{\bibfnamefont{J.-P.}
  \bibnamefont{Poizat}}, \bibinfo{journal}{Nature}
  \textbf{\bibinfo{volume}{396}}, \bibinfo{pages}{537} (\bibinfo{year}{1998}).

\bibitem[{\citenamefont{Steane}(1996)}]{Steane1996}
\bibinfo{author}{\bibfnamefont{A.~M.} \bibnamefont{Steane}},
  \bibinfo{journal}{Phys. Rev. Lett.} \textbf{\bibinfo{volume}{77}},
  \bibinfo{pages}{793} (\bibinfo{year}{1996}).

\bibitem[{\citenamefont{Duer and Briegel}(2003)}]{Duer2003}
\bibinfo{author}{\bibfnamefont{W.}~\bibnamefont{D\"ur}} \bibnamefont{and}
  \bibinfo{author}{\bibfnamefont{H.-J.} \bibnamefont{Briegel}},
  \bibinfo{journal}{Phys. Rev. Lett.} \textbf{\bibinfo{volume}{90}},
  \bibinfo{pages}{067901} (\bibinfo{year}{2003}).

\bibitem[{\citenamefont{Levenson et~al.}(1993)\citenamefont{Levenson, Abram,
  Rivera, Fayolle, Garreau, and Grangier}}]{Levenson1993}
\bibinfo{author}{\bibfnamefont{J.~A.} \bibnamefont{Levenson}},
  \bibinfo{author}{\bibfnamefont{I.}~\bibnamefont{Abram}},
  \bibinfo{author}{\bibfnamefont{T.}~\bibnamefont{Rivera}},
  \bibinfo{author}{\bibfnamefont{P.}~\bibnamefont{Fayolle}},
  \bibinfo{author}{\bibfnamefont{J.~C.} \bibnamefont{Garreau}},
  \bibnamefont{and} \bibinfo{author}{\bibfnamefont{P.}~\bibnamefont{Grangier}},
  \bibinfo{journal}{Phys. Rev. Lett.} \textbf{\bibinfo{volume}{70}},
  \bibinfo{pages}{267} (\bibinfo{year}{1993}).

\bibitem[{\citenamefont{Helmer and Marquardt}(2009)}]{Helmer2009}
\bibinfo{author}{\bibfnamefont{F.}~\bibnamefont{Helmer}} \bibnamefont{and}
  \bibinfo{author}{\bibfnamefont{F.}~\bibnamefont{Marquardt}},
  \bibinfo{journal}{Phys. Rev. A} \textbf{\bibinfo{volume}{79}},
  \bibinfo{pages}{052328} (\bibinfo{year}{2009}).

\bibitem[{\citenamefont{Bishop et~al.}(2009)\citenamefont{Bishop, Tornberg,
  Price, Ginossar, Nunnenkamp, Houck, Gambetta, Koch, Johansson, Girvin
  et~al.}}]{Bishop2009}
\bibinfo{author}{\bibfnamefont{L.~S.} \bibnamefont{Bishop}},
  \bibinfo{author}{\bibfnamefont{L.}~\bibnamefont{Tornberg}},
  \bibinfo{author}{\bibfnamefont{D.}~\bibnamefont{Price}},
  \bibinfo{author}{\bibfnamefont{E.}~\bibnamefont{Ginossar}},
  \bibinfo{author}{\bibfnamefont{A.}~\bibnamefont{Nunnenkamp}},
  \bibinfo{author}{\bibfnamefont{A.~A.} \bibnamefont{Houck}},
  \bibinfo{author}{\bibfnamefont{J.~M.} \bibnamefont{Gambetta}},
  \bibinfo{author}{\bibfnamefont{J.}~\bibnamefont{Koch}},
  \bibinfo{author}{\bibfnamefont{G.}~\bibnamefont{Johansson}},
  \bibinfo{author}{\bibfnamefont{S.~M.} \bibnamefont{Girvin}},
  \bibnamefont{et~al.}, \bibinfo{journal}{New Journal of Physics}
  \textbf{\bibinfo{volume}{11}}, \bibinfo{pages}{073040}
  (\bibinfo{year}{2009}).

\bibitem[{\citenamefont{Wallraff et~al.}(2004)\citenamefont{Wallraff, Schuster,
  Blais, Frunzio, Huang, Majer, Kumar, Girvin, and Schoelkopf}}]{Wallraff2004}
\bibinfo{author}{\bibfnamefont{A.}~\bibnamefont{Wallraff}},
  \bibinfo{author}{\bibfnamefont{D.~I.} \bibnamefont{Schuster}},
  \bibinfo{author}{\bibfnamefont{A.}~\bibnamefont{Blais}},
  \bibinfo{author}{\bibfnamefont{L.}~\bibnamefont{Frunzio}},
  \bibinfo{author}{\bibfnamefont{R.~S.} \bibnamefont{Huang}},
  \bibinfo{author}{\bibfnamefont{J.}~\bibnamefont{Majer}},
  \bibinfo{author}{\bibfnamefont{S.}~\bibnamefont{Kumar}},
  \bibinfo{author}{\bibfnamefont{S.~M.} \bibnamefont{Girvin}},
  \bibnamefont{and} \bibinfo{author}{\bibfnamefont{R.~J.}
  \bibnamefont{Schoelkopf}}, \bibinfo{journal}{Nature (London)}
  \textbf{\bibinfo{volume}{431}}, \bibinfo{pages}{162} (\bibinfo{year}{2004}).

\bibitem[{\citenamefont{Johnson et~al.}(2010)\citenamefont{Johnson, Reed,
  Houck, Schuster, Bishop, Ginossar, Gambetta, DiCarlo, Frunzio, Girvin
  et~al.}}]{Johnson2010}
\bibinfo{author}{\bibfnamefont{B.~R.} \bibnamefont{Johnson}},
  \bibinfo{author}{\bibfnamefont{M.~D.} \bibnamefont{Reed}},
  \bibinfo{author}{\bibfnamefont{A.~A.} \bibnamefont{Houck}},
  \bibinfo{author}{\bibfnamefont{D.~I.} \bibnamefont{Schuster}},
  \bibinfo{author}{\bibfnamefont{L.~S.} \bibnamefont{Bishop}},
  \bibinfo{author}{\bibfnamefont{E.}~\bibnamefont{Ginossar}},
  \bibinfo{author}{\bibfnamefont{J.~M.} \bibnamefont{Gambetta}},
  \bibinfo{author}{\bibfnamefont{L.}~\bibnamefont{DiCarlo}},
  \bibinfo{author}{\bibfnamefont{L.}~\bibnamefont{Frunzio}},
  \bibinfo{author}{\bibfnamefont{S.~M.} \bibnamefont{Girvin}},
  \bibnamefont{et~al.}, \bibinfo{journal}{Nature Physics}
  \textbf{\bibinfo{volume}{6}}, \bibinfo{pages}{663} (\bibinfo{year}{2010}).

\bibitem[{\citenamefont{Boulant et~al.}(2007)\citenamefont{Boulant, Ithier,
  Meeson, Nguyen, Vion, Esteve, Siddiqi, Vijay, Rigetti, Pierre
  et~al.}}]{Boulant2007}
\bibinfo{author}{\bibfnamefont{N.}~\bibnamefont{Boulant}},
  \bibinfo{author}{\bibfnamefont{G.}~\bibnamefont{Ithier}},
  \bibinfo{author}{\bibfnamefont{P.}~\bibnamefont{Meeson}},
  \bibinfo{author}{\bibfnamefont{F.}~\bibnamefont{Nguyen}},
  \bibinfo{author}{\bibfnamefont{D.}~\bibnamefont{Vion}},
  \bibinfo{author}{\bibfnamefont{D.}~\bibnamefont{Esteve}},
  \bibinfo{author}{\bibfnamefont{I.}~\bibnamefont{Siddiqi}},
  \bibinfo{author}{\bibfnamefont{R.}~\bibnamefont{Vijay}},
  \bibinfo{author}{\bibfnamefont{C.}~\bibnamefont{Rigetti}},
  \bibinfo{author}{\bibfnamefont{F.}~\bibnamefont{Pierre}},
  \bibnamefont{et~al.}, \bibinfo{journal}{Phys. Rev. B}
  \textbf{\bibinfo{volume}{76}}, \bibinfo{pages}{014525}
  (\bibinfo{year}{2007}).

\bibitem[{\citenamefont{Mallet et~al.}(2009)\citenamefont{Mallet, Ong,
  Palacios-Laloy, Nguyen, Bertet, Vion, and Esteve}}]{Mallet2009}
\bibinfo{author}{\bibfnamefont{F.}~\bibnamefont{Mallet}},
  \bibinfo{author}{\bibfnamefont{F.~R.} \bibnamefont{Ong}},
  \bibinfo{author}{\bibfnamefont{A.}~\bibnamefont{Palacios-Laloy}},
  \bibinfo{author}{\bibfnamefont{F.}~\bibnamefont{Nguyen}},
  \bibinfo{author}{\bibfnamefont{P.}~\bibnamefont{Bertet}},
  \bibinfo{author}{\bibfnamefont{D.}~\bibnamefont{Vion}}, \bibnamefont{and}
  \bibinfo{author}{\bibfnamefont{D.}~\bibnamefont{Esteve}},
  \bibinfo{journal}{Nature Physics} \textbf{\bibinfo{volume}{5}},
  \bibinfo{pages}{791} (\bibinfo{year}{2009}).

\bibitem[{\citenamefont{Picot et~al.}(2008)\citenamefont{Picot, Lupascu, Saito,
  Harmans, and Mooij}}]{Picot2008}
\bibinfo{author}{\bibfnamefont{T.}~\bibnamefont{Picot}},
  \bibinfo{author}{\bibfnamefont{A.}~\bibnamefont{Lupascu}},
  \bibinfo{author}{\bibfnamefont{S.}~\bibnamefont{Saito}},
  \bibinfo{author}{\bibfnamefont{C.~J. P.~M.} \bibnamefont{Harmans}},
  \bibnamefont{and} \bibinfo{author}{\bibfnamefont{J.~E.} \bibnamefont{Mooij}},
  \bibinfo{journal}{Phys. Rev. B} \textbf{\bibinfo{volume}{78}},
  \bibinfo{pages}{132508} (\bibinfo{year}{2008}).

\bibitem[{\citenamefont{Orlando et~al.}(1999)\citenamefont{Orlando, Mooij,
  Tian, van~der Wal, Levitov, Lloyd, and Mazo}}]{Orlando1999}
\bibinfo{author}{\bibfnamefont{T.~P.} \bibnamefont{Orlando}},
  \bibinfo{author}{\bibfnamefont{J.~E.} \bibnamefont{Mooij}},
  \bibinfo{author}{\bibfnamefont{L.}~\bibnamefont{Tian}},
  \bibinfo{author}{\bibfnamefont{C.~H.} \bibnamefont{van~der Wal}},
  \bibinfo{author}{\bibfnamefont{L.~S.} \bibnamefont{Levitov}},
  \bibinfo{author}{\bibfnamefont{S.}~\bibnamefont{Lloyd}}, \bibnamefont{and}
  \bibinfo{author}{\bibfnamefont{J.~J.} \bibnamefont{Mazo}},
  \bibinfo{journal}{Phys. Rev. B} \textbf{\bibinfo{volume}{60}},
  \bibinfo{pages}{15398} (\bibinfo{year}{1999}).

\bibitem[{\citenamefont{Mooij et~al.}(1999)\citenamefont{Mooij, Orlando,
  Levitov, Tian, van~der Wal, and Lloyd}}]{Mooij1999}
\bibinfo{author}{\bibfnamefont{J.~E.} \bibnamefont{Mooij}},
  \bibinfo{author}{\bibfnamefont{T.~P.} \bibnamefont{Orlando}},
  \bibinfo{author}{\bibfnamefont{L.}~\bibnamefont{Levitov}},
  \bibinfo{author}{\bibfnamefont{L.}~\bibnamefont{Tian}},
  \bibinfo{author}{\bibfnamefont{C.~H.} \bibnamefont{van~der Wal}},
  \bibnamefont{and} \bibinfo{author}{\bibfnamefont{S.}~\bibnamefont{Lloyd}},
  \bibinfo{journal}{Science} \textbf{\bibinfo{volume}{285}},
  \bibinfo{pages}{1036} (\bibinfo{year}{1999}).

\bibitem[{\citenamefont{Paauw et~al.}(2009)\citenamefont{Paauw, Fedorov,
  Harmans, and Mooij}}]{Paauw2009}
\bibinfo{author}{\bibfnamefont{F.~G.} \bibnamefont{Paauw}},
  \bibinfo{author}{\bibfnamefont{A.}~\bibnamefont{Fedorov}},
  \bibinfo{author}{\bibfnamefont{C.~J. P.~M.} \bibnamefont{Harmans}},
  \bibnamefont{and} \bibinfo{author}{\bibfnamefont{J.~E.} \bibnamefont{Mooij}},
  \bibinfo{journal}{Phys. Rev. Lett.} \textbf{\bibinfo{volume}{102}},
  \bibinfo{pages}{090501} (\bibinfo{year}{2009}).

\bibitem[{\citenamefont{Fedorov et~al.}(2010)\citenamefont{Fedorov, Feofanov,
  Macha, Forn-Diaz, Harmans, and Mooij}}]{Fedorov2010}
\bibinfo{author}{\bibfnamefont{A.}~\bibnamefont{Fedorov}},
  \bibinfo{author}{\bibfnamefont{A.~K.} \bibnamefont{Feofanov}},
  \bibinfo{author}{\bibfnamefont{P.}~\bibnamefont{Macha}},
  \bibinfo{author}{\bibfnamefont{P.}~\bibnamefont{Forn-Diaz}},
  \bibinfo{author}{\bibfnamefont{C.~J. P.~M.} \bibnamefont{Harmans}},
  \bibnamefont{and} \bibinfo{author}{\bibfnamefont{J.~E.} \bibnamefont{Mooij}},
  \bibinfo{journal}{Phys. Rev. Lett.} \textbf{\bibinfo{volume}{105}},
  \bibinfo{pages}{060503} (\bibinfo{year}{2010}).

\bibitem[{\citenamefont{Zhu et~al.}(2010)\citenamefont{Zhu, Kemp, Saito, and
  Semba}}]{Zhu2010}
\bibinfo{author}{\bibfnamefont{X.}~\bibnamefont{Zhu}},
  \bibinfo{author}{\bibfnamefont{A.}~\bibnamefont{Kemp}},
  \bibinfo{author}{\bibfnamefont{S.}~\bibnamefont{Saito}}, \bibnamefont{and}
  \bibinfo{author}{\bibfnamefont{K.}~\bibnamefont{Semba}},
  \bibinfo{journal}{Appl. Phys. Lett.} \textbf{\bibinfo{volume}{97}},
  \bibinfo{pages}{102503} (\bibinfo{year}{2010}).

\bibitem[{\citenamefont{Wang et~al.}(2010)\citenamefont{Wang, Chesi, Loss, and
  Bruder}}]{Wang2010}
\bibinfo{author}{\bibfnamefont{Y.-D.} \bibnamefont{Wang}},
  \bibinfo{author}{\bibfnamefont{S.}~\bibnamefont{Chesi}},
  \bibinfo{author}{\bibfnamefont{D.}~\bibnamefont{Loss}}, \bibnamefont{and}
  \bibinfo{author}{\bibfnamefont{C.}~\bibnamefont{Bruder}},
  \bibinfo{journal}{Phys. Rev. B} \textbf{\bibinfo{volume}{81}},
  \bibinfo{pages}{104524} (\bibinfo{year}{2010}).

\bibitem[{\citenamefont{Siddiqi et~al.}(2004)\citenamefont{Siddiqi, Vijay,
  Pierre, M.Wilson, Metcalfe, Rigetti, Frunzio, and Devoret}}]{Siddiqi2004}
\bibinfo{author}{\bibfnamefont{I.}~\bibnamefont{Siddiqi}},
  \bibinfo{author}{\bibfnamefont{R.}~\bibnamefont{Vijay}},
  \bibinfo{author}{\bibfnamefont{F.}~\bibnamefont{Pierre}},
  \bibinfo{author}{\bibfnamefont{C.~M.}~\bibnamefont{Wilson}},
  \bibinfo{author}{\bibfnamefont{M.}~\bibnamefont{Metcalfe}},
  \bibinfo{author}{\bibfnamefont{C.}~\bibnamefont{Rigetti}},
  \bibinfo{author}{\bibfnamefont{L.}~\bibnamefont{Frunzio}}, \bibnamefont{and}
  \bibinfo{author}{\bibfnamefont{M.~H.} \bibnamefont{Devoret}},
  \bibinfo{journal}{Phys. Rev. Lett.} \textbf{\bibinfo{volume}{93}},
  \bibinfo{pages}{207002} (\bibinfo{year}{2004}).

\bibitem[{\citenamefont{Maassen van den Brink
      et~al.}(2005)\citenamefont{Maassen van den Brink, Berkley, and
  Yalowsky}}]{Maassen2005}
\bibinfo{author}{\bibfnamefont{A.} \bibnamefont{Maassen van den Brink}},
  \bibinfo{author}{\bibfnamefont{A.~J.} \bibnamefont{Berkley}},
  \bibnamefont{and} \bibinfo{author}{\bibfnamefont{M.}~\bibnamefont{Yalowsky}},
  \bibinfo{journal}{New Journal of Physics} \textbf{\bibinfo{volume}{7}},
  \bibinfo{pages}{230} (\bibinfo{year}{2005}).

\bibitem[{\citenamefont{Averin and Bruder}(2003)}]{Averin2003}
\bibinfo{author}{\bibfnamefont{D.~V.} \bibnamefont{Averin}} \bibnamefont{and}
  \bibinfo{author}{\bibfnamefont{C.}~\bibnamefont{Bruder}},
  \bibinfo{journal}{Phys. Rev. Lett.} \textbf{\bibinfo{volume}{91}},
  \bibinfo{pages}{057003} (\bibinfo{year}{2003}).

\bibitem[{\citenamefont{Allman et~al.}(2010)\citenamefont{Allman, Altomare,
  Whittaker, Cicak, Li, Sirois, Strong, Teufel, and Simmonds}}]{Allman2010}
\bibinfo{author}{\bibfnamefont{M.~S.} \bibnamefont{Allman}},
  \bibinfo{author}{\bibfnamefont{F.}~\bibnamefont{Altomare}},
  \bibinfo{author}{\bibfnamefont{J.~D.} \bibnamefont{Whittaker}},
  \bibinfo{author}{\bibfnamefont{K.}~\bibnamefont{Cicak}},
  \bibinfo{author}{\bibfnamefont{D.}~\bibnamefont{Li}},
  \bibinfo{author}{\bibfnamefont{A.}~\bibnamefont{Sirois}},
  \bibinfo{author}{\bibfnamefont{J.}~\bibnamefont{Strong}},
  \bibinfo{author}{\bibfnamefont{J.~D.} \bibnamefont{Teufel}},
  \bibnamefont{and} \bibinfo{author}{\bibfnamefont{R.~W.}
  \bibnamefont{Simmonds}}, \bibinfo{journal}{Phys. Rev. Lett.}
  \textbf{\bibinfo{volume}{104}}, \bibinfo{pages}{177004}
  (\bibinfo{year}{2010}).

\bibitem[{\citenamefont{Wang et~al.}(2009)\citenamefont{Wang, Kemp, and
  Semba}}]{Wang2009}
\bibinfo{author}{\bibfnamefont{Y.~D.} \bibnamefont{Wang}},
  \bibinfo{author}{\bibfnamefont{A.}~\bibnamefont{Kemp}}, \bibnamefont{and}
  \bibinfo{author}{\bibfnamefont{K.}~\bibnamefont{Semba}},
  \bibinfo{journal}{Phys. Rev. B} \textbf{\bibinfo{volume}{79}},
  \bibinfo{pages}{024502} (\bibinfo{year}{2009}).

\bibitem[{\citenamefont{Landau and Lifshitz}(1976)}]{Landau_mechanics}
\bibinfo{author}{\bibfnamefont{L.~D.} \bibnamefont{Landau}} \bibnamefont{and}
  \bibinfo{author}{\bibfnamefont{E.~M.} \bibnamefont{Lifshitz}},
  \emph{\bibinfo{title}{Mechanics}}
  (\bibinfo{publisher}{Butterworth-Heinemann}, \bibinfo{year}{1976}).

\bibitem[{\citenamefont{Dykman}(2007)}]{Dykman2007}
\bibinfo{author}{\bibfnamefont{M.~I.}~\bibnamefont{Dykman}},
  \bibinfo{journal}{Phys. Rev. E} \textbf{\bibinfo{volume}{75}},
  \bibinfo{pages}{011101} (\bibinfo{year}{2007}).

\bibitem[{\citenamefont{Serban et~al.}(2010)\citenamefont{Serban, Dykman, and
  Wilhelm}}]{Serban2010}
\bibinfo{author}{\bibfnamefont{I.}~\bibnamefont{Serban}},
  \bibinfo{author}{\bibfnamefont{M.~I.} \bibnamefont{Dykman}},
  \bibnamefont{and} \bibinfo{author}{\bibfnamefont{F.~K.}
  \bibnamefont{Wilhelm}}, \bibinfo{journal}{Phys. Rev. A}
  \textbf{\bibinfo{volume}{81}}, \bibinfo{pages}{022305}
  (\bibinfo{year}{2010}).

\bibitem[{\citenamefont{Lupascu et~al.}(2009)\citenamefont{Lupascu, Bertet,
  Driessen, Harmans, and Mooij}}]{Lupascu2009}
\bibinfo{author}{\bibfnamefont{A.}~\bibnamefont{Lupascu}},
  \bibinfo{author}{\bibfnamefont{P.}~\bibnamefont{Bertet}},
  \bibinfo{author}{\bibfnamefont{E.~F.~C.}~\bibnamefont{Driessen}},
  \bibinfo{author}{\bibfnamefont{C.~J.~P.~M.}~\bibnamefont{Harmans}}, \bibnamefont{and}
  \bibinfo{author}{\bibfnamefont{J.~E.}~\bibnamefont{Mooij}},
  \bibinfo{journal}{Phys. Rev. B} \textbf{\bibinfo{volume}{80}},
  \bibinfo{pages}{172506} (\bibinfo{year}{2009}).

\bibitem[{\citenamefont{Kemp et~al.}(2010)\citenamefont{Kemp, Saito, Munro,
  Nemoto, and Semba}}]{Kemp2010}
\bibinfo{author}{\bibfnamefont{A.}~\bibnamefont{Kemp}},
  \bibinfo{author}{\bibfnamefont{S.}~\bibnamefont{Saito}},
  \bibinfo{author}{\bibfnamefont{W.~J.} \bibnamefont{Munro}},
  \bibinfo{author}{\bibfnamefont{K.}~\bibnamefont{Nemoto}}, \bibnamefont{and}
  \bibinfo{author}{\bibfnamefont{K.}~\bibnamefont{Semba}},
  \bibinfo{journal}{arXiv:1008.4212}  (\bibinfo{year}{2010}).

\end{thebibliography}

\end{document}